\documentclass{fundam}

\usepackage[latin1]{inputenc}
\usepackage{amssymb}

\newtheorem{The}{Theorem}[section]
\newtheorem{Pro}[The]{Proposition}
\newtheorem{Deff}[The]{Definition}
\newtheorem{Lem}[The]{Lemma}

\newcommand{\fa}{\forall}
\newcommand{\Ga}{\Gamma}
\newcommand{\Gas}{\Gamma^\star}
\newcommand{\Gao}{\Gamma^\omega}

\newcommand{\Si}{\Sigma}
\newcommand{\Sis}{\Sigma^\star}
\newcommand{\Sio}{\Sigma^\omega}
\newcommand{\ra}{\rightarrow}
\newcommand{\hs}{\hspace{12mm}

\noi}
\newcommand{\lra}{\leftrightarrow}

\newcommand{\ite}{\item}

\newcommand{\ol}{ $\omega$-language}

\newcommand{\om}{\omega}
\newcommand{\nl}{\newline}
\newcommand{\noi}{\noindent}

\newcommand{\de}{deterministic }
\newcommand{\proo}{\noi {\bf Proof.} }
\newcommand {\ep}{\hfill $\square$}

\begin{document}

\setcounter{page}{1}
\issue{(2008)}

\title{Highly Undecidable Problems about \\ Recognizability by Tiling Systems  }

\address{E Mail: Olivier.Finkel@ens-lyon.fr }

\author{Olivier Finkel\\ {\it Equipe Modèles de Calcul et Complexité}  
 \\ {\it Laboratoire de l'Informatique du Parallélisme}
 \\  CNRS et Ecole Normale Supérieure de Lyon
 \\ 46, Allée d'Italie 69364 Lyon Cedex 07, France.\\ Olivier.Finkel@ens-lyon.fr }
\maketitle

\runninghead{Olivier Finkel}{Highly Undecidable Problems about Recognizability by Tiling Systems }

\begin{abstract}
\noi  Altenbernd, Thomas and 
W\"ohrle have considered  acceptance of 
languages of infinite two-dimensional words (infinite pictures) by finite tiling systems,  
with  usual acceptance conditions, such as  the 
B\"uchi and Muller ones, in  \cite{ATW02}. 
It was proved in \cite{Fin04} 
that it is undecidable whether a B\"uchi-recognizable language of infinite pictures is E-recognizable (respectively, A-recognizable). 
We show here that these two decision problems are actually  $\Pi_2^1$-complete, 
hence located  at the second level of the analytical hierarchy, and ``highly undecidable". 
We give the exact degree of numerous other  undecidable problems for 
B\"uchi-recognizable languages of infinite pictures. In particular, the non-emptiness and the infiniteness problems are $\Si_1^1$-complete, and the  universality 
problem, the inclusion problem, the equivalence problem, the determinizability problem, the complementability problem,  are all $\Pi_2^1$-complete. 
It is also $\Pi_2^1$-complete to determine whether a given B\"uchi 
recognizable language of infinite pictures can be accepted row by row using an automaton model over ordinal words of length 
$\om^2$.  

\end{abstract}


\section{Introduction}
\noi
Languages of infinite words accepted by finite automata were first studied by B\"uchi 
to prove the decidability of the monadic second order theory of one successor
over the integers.  Since then regular $\om$-languages have been much studied and  many applications have been  found  for specification and verification 
of non-terminating systems, 
see \cite{Thomas90,Staiger97,PerrinPin} for many results and references. 
\nl In a recent paper, Altenbernd, Thomas and 
W\"ohrle have considered acceptance of 
languages of infinite two-dimensional words (infinite pictures) by finite tiling systems,  
with the usual acceptance conditions, such as  the 
B\"uchi and Muller ones,  firstly  used for infinite words.
This way they extended both  the classical theory of  $\om$-regular languages and the classical theory of recognizable languages of finite pictures, 
  \cite{Giammarresi-Restivo},  to the case of infinite pictures.  

\hs Many classical decision problems are studied in formal language theory and in automata theory and arise now naturally about 
recognizable languages of infinite pictures. We proved in \cite{Fin04} that many decision problems for B\"uchi-recognizable languages of infinite pictures are 
undecidable. In particular, we showed, using topological arguments,  that 
 it is undecidable whether a B\"uchi-recognizable language of infinite pictures is E-recognizable (respectively, A-recognizable), giving   
  the  answer to two questions raised in  \cite{ATW02}. 
We proved  also several other undecidability results as the following ones: one cannot decide whether a B\"uchi-recognizable language of infinite pictures
can be recognized by a {\it deterministic } B\"uchi or Muller tiling system, or whether it can be accepted row by row 
using an automaton model over ordinal words of length $\om^2$.  

\hs Using   the  $\Pi_2^1$-completeness of the universality problem for $\om$-languages of non deterministic Turing machines which 
was proved by Castro and Cucker in \cite{cc}, and  some  topological arguments, we show in this paper that 
the above decision problems are actually  $\Pi_2^1$-complete, 
hence located  at the second level of the analytical hierarchy, and ``highly undecidable". 
Using other results of \cite{cc}, we give also the exact degree of numerous other  undecidable problems for B\"uchi-recognizable 
languages of infinite pictures. In particular, the non-emptiness and the infiniteness problems are $\Si_1^1$-complete, and the  universality 
problem, the inclusion problem, the equivalence problem,  the complementability problem,  are all $\Pi_2^1$-complete.  
This gives new natural examples of decision problems located at the first or at the second level of the analytical hierarchy.  
We show also that topological properties of B\"uchi-recognizable languages of infinite pictures
  are highly undecidable. 

\hs  The paper is organized as follows. In Section 2 we recall   definitions for 
pictures and tiling systems. The definition and properties of the analytical hierarchy 
are introduced in Section 3.   We recall in Section 4 some notions of topology, including the definitions of Borel and analytic sets. 
We prove high undecidability results  in Section 5.  
Concluding remarks are given in Section 6.

\section{Tiling Systems}

\noi We assume   the reader to be familiar with the theory of formal ($\om$)-languages  
\cite{Thomas90,Staiger97}.
We recall usual notations of formal language theory. 
\nl  When $\Si$ is a finite alphabet, a {\it non-empty finite word} over $\Si$ is any 
sequence $x=a_1\ldots a_k$, where $a_i\in\Sigma$ 
for $i=1,\ldots ,k$ , and  $k$ is an integer $\geq 1$. The {\it length}
 of $x$ is $k$, denoted by $|x|$.
 The {\it empty word} has no letter and is denoted by $\lambda$; its length is $0$. 
 $\Sis$  is the {\it set of finite words} (including the empty word) over $\Sigma$.
 \nl  The {\it first infinite ordinal} is $\om$.
 An $\om$-{\it word} over $\Si$ is an $\om$ -sequence $a_1 \ldots a_n \ldots$, where for all 
integers $ i\geq 1$, ~
$a_i \in\Sigma$.  When $\sigma$ is an $\om$-word over $\Si$, we write
 $\sigma =\sigma(1)\sigma(2)\ldots \sigma(n) \ldots $,  where for all $i$,~ $\sigma(i)\in \Si$,
and $\sigma[n]=\sigma(1)\sigma(2)\ldots \sigma(n)$  for all $n\geq 1$ and $\sigma[0]=\lambda$.
\nl 
 The usual concatenation product of two finite words $u$ and $v$ is 
denoted $u.v$ (and sometimes just $uv$). This product is extended to the product of a 
finite word $u$ and an $\om$-word $v$: the infinite word $u.v$ is then the $\om$-word such that:
\nl $(u.v)(k)=u(k)$  if $k\leq |u|$ , and 
 $(u.v)(k)=v(k-|u|)$  if $k>|u|$.
\nl   
 The {\it set of } $\om$-{\it words } over  the alphabet $\Si$ is denoted by $\Si^\om$.
An  $\om$-{\it language} over an alphabet $\Sigma$ is a subset of  $\Si^\om$.  

\hs We now  define two-dimensional words, i.e. pictures.
\nl  Let $\Si$ be a finite alphabet and $\#$ be a letter not in $\Si$ and let 
$\hat{\Si}=\Si \cup \{\#\}$. If $m$ and $n$ are two integers $>0$ or if $m=n=0$,  
a   picture of size $(m, n)$ over $\Si$ 
is a function $p$ from $\{0, 1, \ldots , m+1\} \times \{0, 1, \ldots , n+1\}$ 
into $\hat{\Si}$ such that 
$p(0, i)=p(m+1, i)=\#$ for all integers $i\in \{0, 1, \ldots , n+1\}$ and 
$p(i, 0)=p(i, n+1)=\#$ for all integers $i\in \{0, 1, \ldots , m+1\}$ and 
$p(i, j) \in \Si$ if $i \notin \{0, m+1\}$ and $j \notin \{0, n+1\}$. The empty picture 
is the only picture of size $(0, 0)$ and is denoted by $\lambda$. Pictures of  size 
$(n, 0)$ or $(0, n)$, for $n>0$, are not defined. $\Si^{\star, \star}$ is the set of 
pictures over $\Si$. A picture language $L$ is a subset of $\Si^{\star, \star}$. 

\hs An $\om$-picture over $\Si$ 
is a function $p$ from $\om \times \om$ into $\hat{\Si}$ such that $p(i, 0)=p(0, i)=\#$ 
for all $i\geq 0$ and $p(i, j) \in \Si$ for $i, j >0$. $\Si^{\om, \om}$ is the set of 
$\om$-pictures over $\Si$. An $\om$-picture language $L$ is a subset of $\Si^{\om, \om}$. 
\nl   For $\Si$ a finite alphabet we call  $\Si^{\om^2}$  the set of functions 
from $\om \times \om$ into $\Si$. So the set $\Si^{\om, \om}$ of $\om$-pictures over 
$\Si$ is a strict subset of $\hat{\Si}^{\om^2}$.  

\hs We shall say that, for each integer $j\geq 1$,  the $j^{th}$ row of an $\om$-picture 
$p\in \Si^{\om, \om}$ is the infinite word $p(1, j).p(2, j).p(3, j) \ldots$ over $\Si$ 
and the $j^{th}$ column of $p$ is the infinite word $p(j, 1).p(j, 2).p(j, 3) \ldots$ 
over $\Si$. 
\nl As usual,  one can imagine that,  for integers $j > k \geq 1$, the $j^{th}$ column of $p$ 
is on the right of the $k^{th}$ column of $p$ and that the $j^{th}$ row of $p$ is 
``above" the $k^{th}$ row of $p$. 

\hs We introduce now tiling systems as in the paper \cite{ATW02}. 
\nl A tiling system is a tuple $\mathcal{A}$=$(Q, \Si, \Delta)$, where $Q$ is a finite set 
of states, $\Si$ is a finite alphabet, $\Delta \subseteq (\hat{\Si} \times Q)^4$ is a finite set 
of tiles. 
\nl A B\"uchi tiling system is a pair $(\mathcal{A},$$ F)$ 
 where $\mathcal{A}$=$(Q, \Si, \Delta)$ 
is a tiling system and $F\subseteq Q$ is the set of accepting states. 
\nl A Muller tiling system is a pair $(\mathcal{A}, \mathcal{F})$ 
where $\mathcal{A}$=$(Q, \Si, \Delta)$ 
is a tiling system and $\mathcal{F}$$\subseteq 2^Q$ is the set of accepting sets of states.

\hs  Tiles are denoted by 
$
\left ( \begin{array}{cc}  (a_3, q_3) & (a_4, q_4)
\\ (a_1, q_1) & (a_2, q_2) \end{array} \right ) \mbox{ with } a_i \in \hat{\Si} \mbox{ and } 
q_i \in Q, 
$

\hs and in general, over an alphabet $\Ga$, by 
$
\left ( \begin{array}{cc} b_3 & b_4 
\\ b_1 & b_2 \end{array} \right ) \mbox{  ~~~~~~with  }  b_i \in \Ga . 
$

\noi A combination of tiles is defined  by: 
\begin{displaymath}
\left ( \begin{array}{cc}  b_3 & b_4
\\ b_1 & b_2 \end{array} \right ) \circ  
\left ( \begin{array}{cc} b'_3 & b'_4
\\ b'_1 & b'_2  \end{array} \right ) = 
\left ( \begin{array}{cc} (b_3, b'_3) & (b_4, b'_4) 
\\ (b_1, b'_1) & (b_2, b'_2) \end{array} \right ) 
\end{displaymath}

\noi A run of a tiling system $\mathcal{A}$=$(Q, \Si, \Delta)$ over a (finite) 
picture $p$ of size $(m, n)$ over $\Si$ 
is a mapping $\rho$  from $\{0, 1, \ldots , m+1\} \times \{0, 1, \ldots , n+1\}$ 
into $Q$ such that for all $(i, j) \in \{0, 1, \ldots , m\} \times \{0, 1, \ldots , n\}$ 
with $p(i, j)=a_{i, j}$ and $\rho(i, j)=q_{i, j}$ we have 
\begin{displaymath}
\left ( \begin{array}{cc}  a_{i, j+1} & a_{i+1, j+1} 
\\  a_{i, j}  & a_{i+1, j}   \end{array} \right ) \circ  
\left ( \begin{array}{cc} q_{i, j+1} & q_{i+1, j+1} 
\\ q_{i, j}  & q_{i+1, j}   \end{array} \right ) \in \Delta . 
\end{displaymath}

\noi A run of a tiling system $\mathcal{A}$=$(Q, \Si, \Delta)$ over an 
$\om$-picture $p \in \Si^{\om, \om}$ 
is a mapping $\rho$  from $\om \times \om$ 
into $Q$ such that for all $(i, j) \in \om \times \om$ 
with $p(i, j)=a_{i, j}$ and $\rho(i, j)=q_{i, j}$ we have 
\begin{displaymath}
\left ( \begin{array}{cc}  a_{i, j+1} & a_{i+1, j+1} 
\\  a_{i, j}  & a_{i+1, j}   \end{array} \right ) \circ  
\left ( \begin{array}{cc} q_{i, j+1} & q_{i+1, j+1} 
\\ q_{i, j}  & q_{i+1, j}   \end{array} \right ) \in \Delta . 
\end{displaymath}

\noi We  now recall  acceptance of finite or  infinite pictures by tiling systems:

\begin{Deff} Let $\mathcal{A}$=$(Q, \Si, \Delta)$ 
be a tiling system, $F\subseteq Q$ and $\mathcal{F}$$\subseteq 2^Q$. 
\begin{itemize}

\ite
 The picture language recognized by $\mathcal{A}$ 
is the set of pictures $p \in \Si^{\star, \star}$ such that there is some run $\rho$ of 
$\mathcal{A}$ on $p$. 

\ite The $\om$-picture language A-recognized  (respectively, E-recognized, B\"uchi-recognized) 
by 
$(\mathcal{A},$$ F)$ 
is the set of $\om$-pictures $p \in \Si^{\om, \om}$ such that there is some run $\rho$ of 
$\mathcal{A}$ on $p$ and $\rho(v) \in F$ for all  
(respectively, for at least one,  for infinitely many) $v\in \om^2$.    It is denoted by $L^A((\mathcal{A},$$ F))$ (respectively,  $L^E((\mathcal{A},$$ F))$, 
$L^B((\mathcal{A},$$ F))$).

\ite The $\om$-picture language Muller-recognized  by $(\mathcal{A}, \mathcal{F})$ is 
the set of $\om$-pictures $p \in \Si^{\om, \om}$ such that there is some run $\rho$ of 
$\mathcal{A}$ on $p$ and $Inf(\rho) \in \mathcal{F}$ where $Inf(\rho)$ is the set of states 
occurring infinitely often in $\rho$. It is denoted by $L^M((\mathcal{A},$$ \mathcal{F}))$.
\end{itemize}
\end{Deff}

\noi Notice that an $\om$-picture language $L \subseteq \Si^{\om, \om}$ is 
recognized by a B\"uchi tiling system if and only if it 
is recognized by a  Muller tiling system, \cite{ATW02}.  
\nl We shall denote $TS(\Si^{\om, \om})$ the class of languages $L \subseteq \Si^{\om, \om}$ 
which are recognized by some  B\"uchi (or Muller) tiling system. 

\hs We recall now an interesting variation of the above defined  acceptance conditions for infinite pictures, introduced in \cite{ATW02}. This 
variation uses the  diagonal   of an $\om$-picture. 

\hs The diagonal of an $\om$-picture $p$ is the set of vertices $Di(p)=\{ (i, i) \mid i\in \om \}$. 

\hs The $\om$-picture language A-recognized  (respectively, E-recognized, B\"uchi-recognized) 
by 
$(\mathcal{A},$$ F)$  {\it  on the diagonal} 
is the set of $\om$-pictures $p \in \Si^{\om, \om}$ such that there is some run $\rho$ of 
$\mathcal{A}$ on $p$ and $\rho(v) \in F$ for all  
(respectively, for at least one,  for infinitely many) $v\in Di(p)$. 
\nl We define similarly the notion of  $\om$-picture language Muller-recognized {\it  on the diagonal} by $(\mathcal{A}, \mathcal{F})$, replacing 
$Inf(\rho)$  by the set of states $Inf(Di(\rho))$ occurring infinitely often  {\it  on the diagonal of}  $\rho$.

\hs The following result was stated  in \cite{ATW02}. 

\begin{The}
An $\om$-picture language $L \subseteq \Si^{\om, \om}$ is 
A-recognized (respectively, E-recognized, B\"uchi-recognized, Muller-recognized)  by a  tiling system if and only if it is 
A-recognized (respectively, E-recognized, B\"uchi-recognized, Muller-recognized) {\it  on the diagonal} by a  tiling system. 
\end{The}

\hs We wish now to see links with classical  notions of tiling of the (quarter of the) plane, see for instance \cite{BJ08-JAC}. 

\hs We denote $\Ga=\hat{\Si} \times Q$ where $\Si$ is the alphabet of pictures and $Q$ is the set of states of a tiling system $\mathcal{A}$=$(Q, \Si, \Delta)$. 
We consider configurations which are elements of $\Ga^{\om \times \om}$. One can imagine that each cell of the quarter of the plane contains a letter 
of the alphabet $\Ga$. 
\nl  Let  $\Delta \subseteq (\hat{\Si} \times Q)^4= \Ga^4$ be  a finite set 
of tiles. We denote its complement   by   $\Delta^-=  \Ga^4 - \Delta$. 
A tiling of  the (quarter of the) plane with $\Delta^-$ as set of  forbidden patterns  is simply a configuration $c\in \Ga^{\om \times \om}$ such that 
for all integers $i, j \in \om$: 
\begin{displaymath}
\left ( \begin{array}{cc}  c(i, j+1) & c(i+1, j+1) 
\\  c(i, j)  & c(i+1, j)   \end{array} \right )   
\in \Delta . 
\end{displaymath}

\noi Then the  $\om$-picture language $L \subseteq \Si^{\om, \om}$ which is A-recognized (respectively, E-recognized, B\"uchi-recognized) {\it on the diagonal}
 by the tiling system $(\mathcal{A},$$ F)$
is simply the set of $\om$-pictures $p \in \Si^{\om, \om}$ which are projections of configurations $c\in \Ga^{\om \times \om}$ which are 
tilings of  the (quarter of the) plane with $\Delta^-$ as set of  forbidden patterns  such that for all 
(respectively, for at least one,  for infinitely many) $i \in \om$  
the second component of  $c(i, i)$ is in $F$. 
A similar characterization can be given for the Muller acceptance condition. 
\nl We can also easily state  similar characterizations for {\it global} recognizability, i.e. not {\it  on the diagonal},  by   tiling systems.

\section{The Analytical Hierarchy}
\noi 
 The set of natural numbers is denoted by $\mathbb{N}$ and the set of all mappings  from $\mathbb{N}$ into $\mathbb{N}$ will 
be denoted by $\mathcal{F}$. 

\hs 
We assume the reader to be familiar with the arithmetical   hierarchy on subsets of  $\mathbb{N}$.   We now recall   the notions
 of analytical hierarchy and of complete sets for classes of this hierarchy which may be found in \cite{rog}.

\begin{Deff}
Let  $k, l >0$ be some  integers. $\Phi$ is a partial recursive function of $k$ function variables and $l$ number variables if there exists $z\in \mathbb{N}$ 
such that for any $(f_1, \ldots , f_k, x_1, \ldots , x_l) \in \mathcal{F}^k \times \mathbb{N}^l$, we have 
$$\Phi (f_1, \ldots , f_k, x_1, \ldots , x_l) = \tau_z^{f_1, \ldots , f_k}(x_1, \ldots , x_l),$$
\noi where the right hand side is the output of the Turing machine with index $z$ and oracles $f_1, \ldots , f_k$ over the input $(x_1, \ldots , x_l)$. 
For $k>0$ and $l=0$, $\Phi$ is a partial recursive function if, for some $z$, 
$$\Phi (f_1, \ldots , f_k) = \tau_z^{f_1, \ldots , f_k}(0).$$
\noi The value $z$ is called the Gödel number or index for $\Phi$. 
\end{Deff}

\begin{Deff}
Let $k, l >0$ be some  integers and $R  \subseteq \mathcal{F}^k \times \mathbb{N}^l$. The relation $R$ is said to be a recursive relation of $k$ 
function variables and $l$ number variables if its characteristic function is recursive. 
\end{Deff}

\noi We now define analytical subsets of $\mathbb{N}^l$.

\begin{Deff}
A subset $R$ of $\mathbb{N}^l$ is analytical if it is recursive or if there exists a recursive set $S  \subseteq \mathcal{F}^m \times \mathbb{N}^n$, with 
$m\geq 0$ and $n\geq l$, such that 
$$R = \{ (x_1, \ldots , x_l) \mid (Q_1s_1)(Q_2s_2) \ldots (Q_{m+n-l}s_{m+n-l}) S(f_1, \ldots , f_m, x_1, \ldots , x_n) \}, $$
\noi where $Q_i$ is either $\fa$ or $\exists$ for $1 \leq i \leq m+n-l$, and where $s_1, \ldots , s_{m+n-l}$ are $f_1, \ldots , f_m, x_{l+1}, \ldots , x_n$ in 
some order. 
\nl The expression $(Q_1s_1)(Q_2s_2) \ldots (Q_{m+n-l}s_{m+n-l}) S(f_1, \ldots , f_m, x_1, \ldots , x_n)$ is called a predicate form for $R$. A
quantifier applying over a function variable is of type $1$, otherwise it is of type $0$. 
In a predicate form the (possibly empty) sequence of quantifiers, indexed by their type, is called the prefix of the form. The reduced prefix is the sequence of 
quantifiers obtained by suppressing the quantifiers of type $0$ from the prefix. 
\end{Deff}

\noi  The levels of the analytical hierarchy are distinguished by considering the number of alternations in the reduced prefix. 

\begin{Deff}
For $n>0$, a $\Si^1_n$-prefix is one whose reduced prefix begins with $\exists^1$ and has $n-1$ alternations of quantifiers. 
A $\Si^1_0$-prefix is one whose reduced prefix is empty. 
For $n>0$, a $\Pi^1_n$-prefix is one whose reduced prefix begins with $\fa^1$ and has $n-1$ alternations of quantifiers. 
A $\Pi^1_0$-prefix is one whose reduced prefix is empty. 
\nl A predicate form is a $\Si^1_n$ ($\Pi^1_n$)-form if it has a  $\Si^1_n$ ($\Pi^1_n$)-prefix. 
The class of sets in some $\mathbb{N}^l$ which can be expressed in $\Si^1_n$-form (respectively, $\Pi^1_n$-form) is denoted by 
$\Si^1_n$   (respectively, $\Pi^1_n$). 
\nl The class $\Si^1_0 = \Pi^1_0$ is the class of arithmetical sets. 
\end{Deff}

\noi We now recall some well known results about the analytical hierarchy. 

\begin{Pro}
Let $R \subseteq \mathbb{N}^l$ for some integer $l$. Then $R$ is an analytical set iff there is some integer $n\geq 0$ such that 
$R \in \Si^1_n$ or $R \in \Pi^1_n$. 
\end{Pro}

\begin{The} For each integer $n\geq 1$, 
\noi 
\begin{enumerate}
\ite[(a)] $\Si^1_n\cup \Pi^1_n \subsetneq  \Si^1_{n +1}\cap \Pi^1_{n +1}$.
\ite[(b)] A set $R \subseteq \mathbb{N}^l$ is in the class $\Si^1_n$ iff its 
complement is in the class $\Pi^1_n$. 
\ite[(c)] $\Si^1_n - \Pi^1_n \neq \emptyset$ and $\Pi^1_n - \Si^1_n \neq \emptyset$.
\end{enumerate}
\end{The}

\noi  Transformations of prefixes  are often used, following the rules given by the next theorem. 

\begin{The}
For any predicate form with the given prefix, an equivalent predicate form with the new one can be obtained, following the 
allowed prefix transformations given below :
\noi 
\begin{enumerate}
\ite[(a)]  $\ldots \exists^0 \exists^0 \ldots \ra \ldots  \exists^0 \ldots, ~~~~~~ \ldots \fa^0  \fa^0 \ldots  \ra \ldots  \fa^0 \ldots ; $
\ite[(b)]  $\ldots \exists^1 \exists^1 \ldots \ra \ldots  \exists^1 \ldots, ~~~~~~ \ldots \fa^1  \fa^1 \ldots  \ra \ldots  \fa^1 \ldots ;$
\ite[(c)]  $\ldots \exists^0 ~~~\ldots  \ra \ldots  \exists^1 \ldots, ~~~~~~ \ldots  \fa^0 ~~~\ldots  \ra \ldots   \fa^1 \ldots ; $
\ite[(d)]  $\ldots \exists^0 \fa^1 \ldots \ra \ldots \fa^1 \exists^0 \ldots, ~~~~ \ldots \fa^0 \exists^1 \ldots  \ra \ldots \exists^1 \fa^0 \ldots ; $
\end{enumerate}
\end{The}

\noi We can now define the notion of 1-reduction and of    $\Si^1_n$-complete (respectively,           $\Pi^1_n$-complete) sets. 
Notice that we give the definition for subsets of  $\mathbb{N}$ but one can  easily extend this definition to the case of  
 subsets of $\mathbb{N}^l$ for some integer $l$. 

\begin{Deff}
Given two sets $A,B \subseteq \mathbb{N}$ we say A is 1-reducible to B and write $A \leq_1 B$
if there exists a total computable injective  function f from      $\mathbb{N}$     to   $\mathbb{N}$     such that  $A = f ^{-1}[B]$. 
\end{Deff}

\begin{Deff}
A set $A \subseteq \mathbb{N}$ is said to be $\Si^1_n$-complete   (respectively,   $\Pi^1_n$-complete)  iff $A$ is a  $\Si^1_n$-set 
 (respectively,   $\Pi^1_n$-set) and for each $\Si^1_n$-set  (respectively,   $\Pi^1_n$-set) $B \subseteq \mathbb{N}$ it holds that 
$B \leq_1 A$. 
\end{Deff}

\noi For each integer $n\geq 1$ there exist some $\Si^1_n$-complete set $E_n \subseteq \mathbb{N}$. The complement $E_n^-=\mathbb{N}-E_n$ is 
a $\Pi^1_n$-complete set. These sets are precisely defined in \cite{rog} or \cite{cc}.

\section{Borel Hierarchy and Analytic Sets}

\noi We assume now the reader to be familiar with basic notions of topology which
may be found in \cite{Moschovakis80,LescowThomas,Kechris94,Staiger97,PerrinPin}.

\hs There is a natural metric on the set $\Sio$ of  infinite words 
over a finite alphabet 
$\Si$ containing at least two letters which is called the {\it prefix metric} and defined as follows. For $u, v \in \Sio$ and 
$u\neq v$ let $\delta(u, v)=2^{-l_{\mathrm{pref}(u,v)}}$ where $l_{\mathrm{pref}(u,v)}$ 
 is the first integer $n$
such that the $(n+1)^{st}$ letter of $u$ is different from the $(n+1)^{st}$ letter of $v$. 
This metric induces on $\Sio$ the usual  Cantor topology for which {\it open subsets} of 
$\Sio$ are in the form $W.\Si^\om$, where $W\subseteq \Sis$.
A set $L\subseteq \Si^\om$ is a {\it closed set} iff its complement $\Si^\om - L$ 
is an open set.
Define now the {\it Borel Hierarchy} of subsets of $\Si^\om$:

\begin{Deff}
For a non-null countable ordinal $\alpha$, the classes ${\bf \Si}^0_\alpha$
 and ${\bf \Pi}^0_\alpha$ of the Borel Hierarchy on the topological space $\Si^\om$ 
are defined as follows:
\nl ${\bf \Si}^0_1$ is the class of open subsets of $\Si^\om$, 
 ${\bf \Pi}^0_1$ is the class of closed subsets of $\Si^\om$, 
\nl and for any countable ordinal $\alpha \geq 2$: 
\nl ${\bf \Si}^0_\alpha$ is the class of countable unions of subsets of $\Si^\om$ in 
$\bigcup_{\gamma <\alpha}{\bf \Pi}^0_\gamma$.
 \nl ${\bf \Pi}^0_\alpha$ is the class of countable intersections of subsets of $\Si^\om$ in 
$\bigcup_{\gamma <\alpha}{\bf \Si}^0_\gamma$.
\end{Deff}

\noi For 
a countable ordinal $\alpha$,  a subset of $\Si^\om$ is a Borel set of {\it rank} $\alpha$ iff 
it is in ${\bf \Si}^0_{\alpha}\cup {\bf \Pi}^0_{\alpha}$ but not in 
$\bigcup_{\gamma <\alpha}({\bf \Si}^0_\gamma \cup {\bf \Pi}^0_\gamma)$.

\hs    
There are also some subsets of $\Si^\om$ which are not Borel. 
Indeed there exists another hierarchy beyond the Borel hierarchy, which is called the 
projective hierarchy and which is obtained from  the Borel hierarchy by 
successive applications of operations of projection and complementation.
The first level of the projective hierarchy is formed by the class of {\it analytic sets} and the class of {\it co-analytic sets} which are complements of 
analytic sets.  
In particular 
the class of Borel subsets of $\Si^\om$ is strictly included into 
the class  ${\bf \Si}^1_1$ of {\it analytic sets} which are 
obtained by projection of Borel sets. 

\begin{Deff} 
A subset $A$ of  $\Si^\om$ is in the class ${\bf \Si}^1_1$ of {\bf analytic} sets
iff there exists another finite set $Y$ and a Borel subset $B$  of  $(\Si \times Y)^\om$ 
such that $ x \in A \lra \exists y \in Y^\om $ such that $(x, y) \in B$, 
where $(x, y)$ is the infinite word over the alphabet $\Si \times Y$ such that
$(x, y)(i)=(x(i),y(i))$ for each  integer $i\geq 1$.
\end{Deff} 
 
\noi  We now define completeness with regard to reduction by continuous functions. 
For a countable ordinal  $\alpha\geq 1$, a set $F\subseteq \Si^\om$ is said to be 
a ${\bf \Si}^0_\alpha$  
(respectively,  ${\bf \Pi}^0_\alpha$, ${\bf \Si}^1_1$)-{\it complete set} 
iff for any set $E\subseteq Y^\om$  (with $Y$ a finite alphabet): 
 $E\in {\bf \Si}^0_\alpha$ (respectively,  $E\in {\bf \Pi}^0_\alpha$,  $E\in {\bf \Si}^1_1$) 
iff there exists a continuous function $f: Y^\om \ra \Si^\om$ such that $E = f^{-1}(F)$. 
 ${\bf \Si}^0_n$
 (respectively ${\bf \Pi}^0_n$)-complete sets, with $n$ an integer $\geq 1$, 
 are thoroughly characterized in \cite{Staiger86a}.  

\hs In particular  $\mathcal{R}=(0^\star.1)^\om$  
is a well known example of 
${\bf \Pi}^0_2 $-complete subset of $\{0, 1\}^\om$. It is the set of 
$\om$-words over $\{0, 1\}$ having infinitely many occurrences of the letter $1$. 
Its  complement 
$\{0, 1\}^\om - (0^\star.1)^\om$ is a 
${\bf \Si}^0_2 $-complete subset of $\{0, 1\}^\om$.

\hs  For $\Ga$ a finite alphabet having at least two letters, the 
set $\Ga^{\om \times \om}$ of functions  from $\om \times \om$ into $\Ga$ 
is usually equipped with the product topology  of the discrete 
topology on $\Ga$. 
This topology may be defined 
by the following distance $d$. Let $x$ and $y$  in $\Ga^{\om  \times \om}$ 
such that $x\neq y$, then  
$$ d(x, y)=\frac{1}{2^n} ~~~~~~~\mbox{   where  }$$
$$n=min\{p\geq 0 \mid  \exists (i, j) ~~ x(i, j)\neq y(i, j) \mbox{ and } i+j=p\}.$$

\noi  Then the topological space $\Ga^{\om \times \om}$ is homeomorphic to the 
topological space $\Ga^{\om}$, equipped with the Cantor topology.  
Borel subsets  of   $\Ga^{\om  \times \om}$ are defined from open 
subsets  as in the case of the topological space $\Ga^\om$. 
Analytic  subsets  of   $\Ga^{\om  \times \om}$ are obtained as projections on 
$\Ga^{\om \times \om}$
 of Borel subsets of the product space  $\Ga^{\om \times \om} \times \Ga^\om$.
\nl  The set $\Si^{\om, \om}$ of $\om$-pictures over $\Si$, 
viewed as a topological subspace of $\hat{\Si}^{\om \times \om}$, 
 is easily seen to be homeomorphic to the topological space $\Si^{\om \times \om}$, 
via the mapping 
$\varphi: \Si^{\om, \om} \ra \Si^{\om \times \om}$ 
defined by $\varphi(p)(i, j)=p(i+1, j+1)$ for all 
$p\in \Si^{\om, \om}$ and $i, j \in \om$.

\section{Highly Undecidable Problems}

\noi  We are now going to study decision problems about recognizable languages of infinite pictures. We shall use some results of Castro and Cucker 
who  studied degrees of decision problems for $\om$-languages accepted by Turing machines and proved that many of them are highly undecidable, \cite{cc}. 
\nl So we now recall  the notion of acceptance of infinite words by Turing machines considered by Castro and Cucker in \cite{cc}. 

\begin{Deff}
A non deterministic Turing machine $\mathcal{M}$ is a $5$-tuple $\mathcal{M}=(Q, \Si, \Ga, \delta, q_0)$, where $Q$ is a finite set of states, 
$\Si$ is a finite input alphabet, $\Ga$ is a finite tape alphabet satisfying $\Si  \subseteq \Ga$, $q_0$ is the initial state, 
and $\delta$ is a mapping from $Q \times \Ga$ to subsets of $Q \times \Ga \times \{L, R, S\}$. A configuration of $\mathcal{M}$ is a $3$-tuple 
$(q, \sigma, i)$, where $q\in Q$, $\sigma \in \Ga^\om$ and $i\in \mathbb{N}$. An infinite sequence of configurations $r=(q_i, \alpha_i, j_i)_{i\geq 1}$
is called a run of $\mathcal{M}$ on $w\in \Sio$ iff: 
\begin{enumerate}
\ite[(a)] $(q_1, \alpha_1, j_1)=(q_0, w, 1)$, and 
\ite[(b)] for each $i\geq 1$, $(q_i, \alpha_i, j_i) \vdash (q_{i+1}, \alpha_{i+1}, j_{i+1})$, 
\end{enumerate}
\noi where $\vdash$ is the transition relation of $\mathcal{M}$ defined as usual. The run $r$ is said to be complete if 
 the limsup of the head positions is infinity, i.e. if $(\fa n \geq 1) (\exists k \geq 1) (j_k \geq n)$. 
The run $r$ is said to be oscillating if the liminf of the head positions is bounded, i.e. if $(\exists k \geq 1) (\fa n \geq 1) (\exists m \geq n) ( j_m=k)$. 

\end{Deff}

\begin{Deff}
Let $\mathcal{M}=(Q, \Si, \Ga, \delta, q_0)$ be a non deterministic Turing machine   and $F \subseteq Q$. The $\om$-language accepted by $(\mathcal{M}, F)$ is 
the set of $\om$-words $ \sigma \in \Sio$ such that there exists a complete non oscillating run $ r=(q_i, \alpha_i, j_i)_{i\geq 1}$
 of $\mathcal{M}$  on  $\sigma$ such that, for all $ i, q_i \in F.$
\end{Deff}

\noi The above acceptance condition is denoted $1'$-acceptance in  \cite{CG78b}. Another usual acceptance condition is the now called B\"uchi 
acceptance condition which is also denoted $2$-acceptance in  \cite{CG78b}. 
We now  recall its definition. 

\begin{Deff}
Let $\mathcal{M}=(Q, \Si, \Ga, \delta, q_0)$ be a non deterministic Turing machine   
and $F \subseteq Q$. The $\om$-language B\"uchi accepted by $(\mathcal{M}, F)$ is 
the set of $\om$-words $ \sigma \in \Sio$ such that there exists a complete non oscillating run $ r=(q_i, \alpha_i, j_i)_{i\geq 1}$
 of $\mathcal{M}$  on  $\sigma$ and  infinitely many integers $i$ such that $q_i \in F.$
\end{Deff}

\noi Recall that Cohen and Gold proved in \cite[Theorem 8.6]{CG78b} that one can effectively construct, from a given non deterministic Turing machine, 
another equivalent non deterministic Turing machine, equipped with the same kind of  acceptance condition, and in which every run is complete non oscillating. 
 Cohen and Gold proved also in \cite[Theorem 8.2]{CG78b} that an $\om$-language is accepted by a non deterministic Turing machine with 
$1'$-acceptance condition iff it is accepted by a non deterministic Turing machine with B\"uchi acceptance condition.

\hs From now on, we shall  denote $\mathcal{M}_z$ the non deterministic Turing machine of index $z$, (accepting words over $\Si=\{a, b\}$), 
equipped with a $1'$-acceptance condition. In a similar way we shall  denote $\mathcal{T}_z$ the non deterministic tiling system of index $z$, 
(accepting pictures over $\Si=\{a, b\}$),
 equipped with a B\"uchi acceptance condition.

\hs  For $\sigma \in \Sio=\{a, b\}^\om$ we denote $\sigma^a$ the $\om$-picture whose first row is the $\om$-word $\sigma$ and whose other rows 
are labelled with the letter $a$. 
 For an \ol~  $L \subseteq \Sio=\{a, b\}^\om$  we  denote  $L^a$  the language of infinite pictures $ \{ \sigma^a \mid \sigma \in L \}$.

\hs We can now recall a result proved in \cite{Fin04} which will be useful later. 

\begin{Lem}[\cite{Fin04}] \label{lemTM}
If   $L \subseteq \Sio$ is 
accepted by some Turing machine (in which every run is complete non oscillating) with a B\"uchi acceptance 
condition, then $L^a$ is B\"uchi recognizable by a finite tiling system. 
\end{Lem}

\proo  
Let $L \subseteq \Sio$ be an \ol~ accepted by some Turing machine $\mathcal{M}=(Q, \Si, \Ga, \delta, q_0)$ 
with a B\"uchi acceptance condition, where $F \subseteq Q$ is the set of accepting states. 
\nl  We assume that the Turing machine has a single semi-infinite tape, 
with one reading head which may 
also write on the tape. 
\nl   Cohen and Gold proved that one can consider only such a restricted model 
of Turing machines \cite{CG78b}. 
\nl An instantaneous configuration of  $\mathcal{M}$ is given by an infinite word 
$u.q.v$ where $u\in \Gas$, $q\in Q$,  $v\in \Gao$, and the first letter of $v$ is the one 
scanned by the head of $\mathcal{M}$. 
\nl The initial configuration of  $\mathcal{M}$ reading the  infinite word $\sigma \in \Sio$ 
is  $q_0.\sigma$.  
\nl A computation of $\mathcal{M}$ reading $\sigma \in \Sio$ is an infinite sequence of 
configurations $\alpha_0, \alpha_1, \alpha_2, \ldots ,  \alpha_i, \ldots $~~, 
where $\alpha_0=q_0.\sigma$ is the  initial configuration and for all integers $i\geq 0$, 
 $\alpha_i=u_i.q_i.v_i$ is the $(i+1)^{th}$ configuration. 
\nl The computation is successful if and only if there exists a final state $q_f \in F$ and 
infinitely many integers $i$ such that $q_i=q_f$. 

\hs Using  a similar reasoning as in the classical 
proof of the undecidability of the emptiness 
problem for recognizable languages of finite pictures,  
 \cite[p. 34]{Giammarresi-Restivo},  
we can define a set of tiles $\Delta$ in  such a way that for $\sigma \in \Sio$, 
a run $\rho$ of the tiling system 
$\mathcal{T}$=$(\Si, \Ga \cup Q, \Delta, F)$ over the infinite picture $\sigma^a$ 
satifies:
$$\mbox{ for each integer } i \geq 0 ~~~~ 
\rho(0, i).\rho(1, i).\rho(2, i)\ldots = \alpha_i=u_i.q_i.v_i$$

\noi i.e. $\rho(0, i).\rho(1, i).\rho(2, i)\ldots $ is the $(i+1)^{th}$ configuration of 
$\mathcal{M}$ reading the $\om$-word $\sigma \in \Sio$. 
\nl Thus the B\"uchi tiling system $(\mathcal{T},$$ F)$ recognizes the language 
$L^a$.  
\ep  

\hs Notice that the above cited constructions of \cite{CG78b} and of the  proof of Lemma \ref{lemTM}
are effective and that they can be achieved in an injective way. 
This is expressed by the following lemma. 

\begin{Lem}\label{red1}
There is an injective computable function 
$K$ from $\mathbb{N}$ into $\mathbb{N}$ satisfying the following property. 
\nl  If $\mathcal{M}_z$ is the non deterministic Turing machine (equipped with a $1'$-acceptance condition) of index $z$, 
and if $\mathcal{T}_{K(z)}$ is the tiling system 
(equipped with a B\"uchi  acceptance condition) of index $K(z)$, 
then  
$$L(\mathcal{M}_z)^a  = L^B(\mathcal{T}_{K(z)})$$ 
\end{Lem}

\noi  Recall that Castro and Cucker proved in \cite{cc} that the non-emptiness problem and the infiniteness problem for $\om$-languages 
of Turing machines are both
$\Si_1^1$-complete. We can now easily infer from Lemma \ref{red1} a similar result for recognizable languages of infinite pictures. 

\begin{The}\label{E-I}
The  non-emptiness  problem and the infiniteness problem for B\"uchi-recognizable languages of infinite pictures 
 are $\Si_1^1$-complete, i.e. : 
\begin{enumerate}
\ite  $\{ z \in \mathbb{N} \mid  L^B(\mathcal{T}_z) \neq  \emptyset  \}$ is  $\Si_1^1$-complete.  
\ite  $\{ z \in \mathbb{N} \mid  L^B(\mathcal{T}_z)  \mbox{ is infinite }   \}$ is  $\Si_1^1$-complete.
\end{enumerate}
\end{The}

\proo We first show  that these two decision problems are in the class $\Si_1^1$. 
\nl Notice first that, using a recursive bijection $b : (\mathbb{N}-\{0\})^2  \ra \mathbb{N}-\{0\}$,  
one can associate to each  $\om$-word $\sigma \in \Sio$ a unique  $\om$-picture $p^\sigma \in \Si^{\om, \om}$ which is simply defined 
by $p^\sigma(i, j)=\sigma(b(i, j))$ for all integers $i, j \geq 1$. 
\nl On the other hand a run of a tiling system  $\mathcal{A}$=$(Q, \Si, \Delta)$ over an 
$\om$-picture $p \in \Si^{\om, \om}$ 
is a mapping $\rho$  from $\om \times \om$ 
into $Q$, i.e. an element of $Q^{\om \times \om}$. Using again a recursive bijection between $(\mathbb{N})^2$ and $\mathbb{N}$,  we 
can identify a run $\rho$ with an element of $Q^\om$ and finally with a coding of this element over the alphabet $\{0, 1\}$. So the run  $\rho$ can be identified 
with its code $\bar{\rho} \in  \{0, 1\}^\om$. 
\nl Assume now that the tiling system  $\mathcal{A}$=$(Q, \Si, \Delta)$ is equipped with a set of  accepting states $F \subseteq Q$. It  is then  easy to see that 
for $\sigma \in \Sio$ and $\rho \in  \{0, 1\}^\om$, ~~  ``$\rho$ is a B\"uchi accepting run of $(\mathcal{A}, F)$ over the $\om$-picture $p^\sigma$"
can be expressed by an arithmetical formula, see also \cite[Section 2.4]{ATW02}. 

\hs We can now express ``$L^B(\mathcal{T}_z) \neq  \emptyset$" by ``$\exists \sigma \in \Sio ~~  \exists \rho \in  \{0, 1\}^\om$ ~~[  $\rho$ is a B\"uchi accepting run of 
$\mathcal{T}_z$ over the $\om$-picture $p^\sigma$ ]"  which is a  $\Si_1^1$-formula. 

\hs In order to show that ``$L^B(\mathcal{T}_z)  \mbox{ is infinite }$" can be also expressed by a  $\Si_1^1$-formula, we shall use again the bijection 
$b : (\mathbb{N}-\{0\})^2  \ra \mathbb{N}-\{0\}$. 
\nl We can consider an infinite word $\sigma \in \Sio$ as a countably infinite family of infinite words over $\Si$ :  the family of $\om$-words 
$(\sigma_i)$ such that for each $i \geq 1$, $\sigma_i$ is defined by $\sigma_i(j)= \sigma(b(i, j))$ for each $j\geq 1$. 
In a similar manner an $\om$-word $\rho \in  \{0, 1\}^\om$ can be considered as a countably infinite family of infinite words 
$(\rho_i)$  defined, for each $i \geq 1$,   by $\rho_i(j)= \rho(b(i, j))$ for each $j\geq 1$. 
\nl We can now express ``$L^B(\mathcal{T}_z)  \mbox{ is infinite }$" by the formula ``$\exists \sigma \in \Sio ~~  \exists \rho \in  \{0, 1\}^\om$ ~~[ 
( all $\om$-words $\sigma_i$ are distinct ) and (for each integer $i \geq 1$,  $\rho_i$ is a B\"uchi accepting run of 
$\mathcal{T}_z$ over the $\om$-picture $p^{\sigma_i}$) ]" . 
This is a $\Si_1^1$-formula because ``all $\om$-words $\sigma_i$ are distinct" can be expressed by the arithmetical formula:
``$(\fa j > k \geq 1)  (\exists i \geq 1) ~ \sigma_j(i) \neq \sigma_k(i)$".

\hs Using the reduction $K$ given by Lemma \ref{red1} we can easily see that $L(\mathcal{M}_z)$ is empty (respectively, infinite) 
if and only if   $L^B(\mathcal{T}_{K(z)})=L(\mathcal{M}_z)^a$ is empty (respectively, infinite). This proves that 
 $$\{ z \in \mathbb{N} \mid  L(\mathcal{M}_z) \neq   \emptyset \} \leq_1 \{ z \in \mathbb{N} \mid  L^B(\mathcal{T}_z) \neq   \emptyset \}$$ 
$$\{ z \in \mathbb{N} \mid  L(\mathcal{M}_z) \mbox{ is infinite }  \} \leq_1 \{ z \in \mathbb{N} \mid  L^B(\mathcal{T}_z) \mbox{ is infinite }  \}$$ 
\noi and then the completeness result follows from the $\Si_1^1$-completeness of the non-emptiness problem 
and of the infiniteness problem for $\om$-languages of Turing machines. 
\ep 

\hs On the other hand it is easy to see that the language $\Si^{\om, \om}- (\Sio)^a$ of $\om$-pictures is B\"uchi recognizable. But the class 
$TS(\Si^{\om, \om})$ is closed under finite union, so  we get the following result. 

\begin{Lem}\label{lem2'}
If   $L \subseteq \Sio$ is 
accepted by some Turing machine with a B\"uchi acceptance 
condition, then $L^a \cup  [ \Si^{\om, \om} - (\Sio)^a ]$ is B\"uchi recognizable by a finite tiling system. 
\end{Lem}

\noi Notice that the constructions are effective and that they can be achieved in an injective way, so we can now state  the following lemma, 
asserting   the existence of a computable function $H$ which will be often used in the sequel.

\begin{Lem}\label{red}
There is an injective computable function 
$H$ from $\mathbb{N}$ into $\mathbb{N}$ satisfying the following property. 
\nl  If $\mathcal{M}_z$ is the non deterministic Turing machine (equipped with a $1'$-acceptance condition) of index $z$, 
and if $\mathcal{T}_{H(z)}$ is the tiling system 
(equipped with a B\"uchi  acceptance condition) of index $H(z)$, 
then  
$$L(\mathcal{M}_z)^a \cup  [ \Si^{\om, \om}- (\Sio)^a ] = L^B(\mathcal{T}_{H(z)})$$ 
\end{Lem}

\hs We can now prove that the universality  problem for B\"uchi-recognizable languages of infinite pictures 
 is highly undecidable and give its exact degree. 
  
\begin{The}\label{U}
The universality  problem for B\"uchi-recognizable languages of infinite pictures 
 is $\Pi_2^1$-complete, i.e. : ~~
 $\{ z \in \mathbb{N} \mid  L^B(\mathcal{T}_z) =  \Si^{\om, \om} \}$ is  $\Pi_2^1$-complete. 
\end{The}

\proo We first check that  the  set $\{ z \in \mathbb{N} \mid  L^B(\mathcal{T}_z) =  \Si^{\om, \om} \}$
is in the class $\Pi_2^1$.  We can  write that 
$L^B(\mathcal{T}_z) = \Si^{\om, \om}$ if and only if  
``$\fa$ $\sigma \in \Si^{\om}$ $\exists \rho \in \{0,1\}^\om $ ~~( $\rho$ is a B\"uchi-accepting run of $\mathcal{T}_z$ over $p^\sigma$ )".
The two quantifiers of type 1 are followed by an arithmetical formula. Thus the set 
$\{ z \in \mathbb{N} \mid  L^B(\mathcal{T}_z) = \Si^{\om, \om} \}$ is in 
the class  $\Pi_2^1$. 

\hs In order to prove completeness we  use the corresponding result for Turing machines proved in \cite{cc}: the set 
$\{ z \in \mathbb{N} \mid  L(\mathcal{M}_z) = \Si^\om \}$ is $\Pi_2^1$-complete. 
Consider now the injective  computable function  $H$ from $\mathbb{N}$ into $\mathbb{N}$ given in Lemma \ref{red}. 
It is easy to see that for any Turing machine $\mathcal{M}_z$ it holds that $ L(\mathcal{M}_z) = \Si^\om$ if and only if 
$L(\mathcal{M}_z)^a \cup  [ \Si^{\om, \om}- (\Sio)^a ] = L^B(\mathcal{T}_{H(z)})= \Si^{\om, \om}$. This proves that 
$\{ z \in \mathbb{N} \mid  L(\mathcal{M}_z) = \Si^\om \} \leq_1 \{ z \in \mathbb{N} \mid  L^B(\mathcal{T}_z) = \Si^{\om, \om} \}$, thus this latter 
set is $\Pi_2^1$-complete. 
\ep

\hs We now consider   
 the inclusion and the equivalence problems for B\"uchi-recognizable languages of infinite pictures. 

\begin{The}\label{inclusion-equivalence}
\noi  The inclusion and the equivalence problems for B\"uchi-recognizable languages of infinite pictures are 
  $\Pi_2^1$-complete, i.e. : 
\begin{enumerate}
\ite $\{ (y, z) \in \mathbb{N}^2  \mid  L^B(\mathcal{T}_y) \subseteq L^B(\mathcal{T}_z)  \}$ is  $\Pi_2^1$-complete. 
\ite $\{ (y, z) \in \mathbb{N}^2  \mid  L^B(\mathcal{T}_y) = L^B(\mathcal{T}_z)  \}$ is  $\Pi_2^1$-complete. 
\end{enumerate}
\end{The}

\proo  We first prove that the set $\{ (y, z) \in \mathbb{N}^2  \mid  L^B(\mathcal{T}_y) \subseteq L^B(\mathcal{T}_z)  \}$
 is a $\Pi_2^1$-set. 
It suffices to remark that  ``$L^B(\mathcal{T}_y) \subseteq L^B(\mathcal{T}_z)$" can be expressed by the  $\Pi_2^1$-formula :
``$\fa$ $\sigma \in \Si^{\om}$ $\fa \rho \in \{0,1\}^\om $ $\exists \rho' \in \{0,1\}^\om$ 
 [if ($\rho$ is a B\"uchi accepting run of $\mathcal{T}_y$ over $p^\sigma$), then ($\rho'$ is a  B\"uchi accepting run of $\mathcal{T}_z$ over $p^\sigma$)]".
\nl Then the set  $\{ (y, z) \in \mathbb{N}^2  \mid     L^B(\mathcal{T}_y) = L^B(\mathcal{T}_z) \}$ which is the intersection of the two sets 
$\{ (y, z) \in \mathbb{N}^2  \mid   L^B(\mathcal{T}_y) \subseteq L^B(\mathcal{T}_z)  \}$ and 
$\{ (y, z) \in \mathbb{N}^2  \mid  L^B(\mathcal{T}_z) \subseteq L^B(\mathcal{T}_y)   \}$ is also a  $\Pi_2^1$-set.

\hs On the other hand it is easy to check that for all integers $y, z$, it holds that 
$L(\mathcal{M}_y)  \subseteq  L(\mathcal{M}_z)$ iff  $L^B(\mathcal{T}_{H(y)}) \subseteq L^B(\mathcal{T}_{H(z)}) $ and that 
$L(\mathcal{M}_y)  =  L(\mathcal{M}_z)$ iff  $L^B(\mathcal{T}_{H(y)}) = L^B(\mathcal{T}_{H(z)}) $. Thus using the reduction $H$ we see that 
$$\{ (y, z) \in \mathbb{N}^2  \mid   L(\mathcal{M}_y)  \subseteq  L(\mathcal{M}_z) \}
\leq_1  \{ (y, z) \in \mathbb{N}^2  \mid  L^B(\mathcal{T}_y) \subseteq L^B(\mathcal{T}_z)  \}$$ 
$$\{ (y, z) \in \mathbb{N}^2  \mid   L(\mathcal{M}_y)  =  L(\mathcal{M}_z) \}
\leq_1  \{ (y, z) \in \mathbb{N}^2  \mid  L^B(\mathcal{T}_y) = L^B(\mathcal{T}_z)  \}$$ 
\noi The $\Pi_2^1$-completeness follows then from the $\Pi_2^1$-completeness of the inclusion and the equivalence problems for 
$\om$-languages of Turing machines proved in \cite{cc}. 
\ep 

\hs We are going to consider now the decision problems studied in \cite{Fin04}. 
Using topological arguments,  we 
gave in \cite{Fin04}  the  answer to two questions raised in  \cite{ATW02}, 
showing that it is undecidable whether a B\"uchi 
recognizable language of infinite pictures is E-recognizable (respectively, A-recognizable). 
We are going to show that these problems are actually $\Pi_2^1$-complete, using again some topological arguments.

\begin{The}\label{EA}
\noi  The  problem to determine whether a given B\"uchi-recognizable language of infinite pictures is E-recognizable (respectively, A-recognizable) 
is  $\Pi_2^1$-complete,  i.e. :  
\begin{enumerate}
\ite $\{  z \in \mathbb{N}  \mid  L^B(\mathcal{T}_z) \mbox{ is  E-recognizable  }\}\mbox{  is } \Pi_2^1\mbox{-complete.}$
\ite  $\{  z \in \mathbb{N}  \mid  L^B(\mathcal{T}_z) \mbox{ is  A-recognizable  }\}\mbox{  is } \Pi_2^1\mbox{-complete.}$
\end{enumerate}
\end{The}

\proo   We first prove that   the set  
$\{  z \in \mathbb{N}  \mid  L^B(\mathcal{T}_z) \mbox{ is  E-recognizable  }\}$  is a  $\Pi_2^1$-set.   The sentence 
``$L^B(\mathcal{T}_z) \mbox{ is  E-recognizable}$"  can be expressed by 
``$\exists y ~~ L^B(\mathcal{T}_z) =  L^E(\mathcal{T}_y) $". 
The assertion  ``$L^B(\mathcal{T}_z) =  L^E(\mathcal{T}_y)$" can be expressed by a $\Pi_2^1$-formula in a very similar manner as 
``$L^B(\mathcal{T}_z) =  L^B(\mathcal{T}_y)$" was (see the  proof of Theorem \ref{inclusion-equivalence}), because 
for  $\sigma \in \Si^{\om}$ and  $\rho \in \{0,1\}^\om$ the sentence 
 ``($\rho$ is a E-accepting run of $\mathcal{T}_z$ over the $\om$-picture $p^\sigma$)" can be expressed by an arithmetical formula. 
 Moreover the  quantifier $\exists y $ is of type $0$ thus ``$L^B(\mathcal{T}_z) \mbox{ is  E-recognizable}$" can be expressed by a 
$\Pi_2^1$-formula.
\nl We prove in a very similar manner that  ``$L^B(\mathcal{T}_z)$ is  A-recognizable" can be expressed by a 
$\Pi_2^1$-formula.
 Details are here left to the reader.

\hs We now prove the completeness part of the result. 
We first define a simple operation over $\om$-languages. For two $\om$-words $x, x' \in \Sio$ the $\om$-word $x \otimes x'$ is just the shuffle of the 
two $\om$-words $x$ and  $x'$ which is simply  defined by : 
for every integer $n\geq 1$ ~$(x \otimes x')(2n -1)=x(n)$ and $(x \otimes x')(2n)=x'(n)$. 
For two $\om$-languages $L, L' \subseteq \Sio$, the $\om$-language $L \otimes L' $ is defined by $L \otimes L' = \{ x \otimes x' \mid x\in L \mbox{ and } 
x'\in L' \}$. 

\hs We shall   use the following construction. 
We  know that there is a simple example of ${\bf \Si}^1_1$-complete set $L \subseteq \Sio$  accepted by a $1$-counter automaton, hence by a Turing machine 
with $1'$ acceptance condition, see \cite{Fin03a}. 
Then it is easy to define an injective computable function $\theta$ from $\mathbb{N}$ into $\mathbb{N}$ such that, for every integer $z \in \mathbb{N}$, 
it holds that $L(\mathcal{M}_{\theta(z)}) = (L  \otimes   \Sio)  \cup     (\Sio    \otimes   L(\mathcal{M}_{z}))$.   

\hs We are going to use now the reduction $H$ already considered above. We have seen that  
 $$L(\mathcal{M}_z) = \Si^\om \mbox{ if and only if }L(\mathcal{T}_{H(z) })= \Si^{\om, \om}$$ 
\noi and we can easily see that 
 $$L(\mathcal{M}_{\theta(z)}) = \Si^\om \mbox{ if and only if } L(\mathcal{M}_z)= \Si^\om$$
\noi because $L \neq \Sio$. 

\hs The reduction $H \circ \theta$  is an injective computable function  from $\mathbb{N}$ into $\mathbb{N}$. 
\nl We consider now two cases. 
\nl {\bf First case.}  $L(\mathcal{M}_z)= \Si^\om$. Then $L(\mathcal{M}_{\theta(z)}) = \Si^\om$ and 
$L^B(\mathcal{T}_{H \circ \theta (z) })= \Si^{\om, \om}$. In particular $L^B(\mathcal{T}_{H \circ \theta (z) })$ is E-recognized (respectively,  A-recognized) 
by a tiling system. 
\nl {\bf Second case.} $L(\mathcal{M}_z) \neq  \Si^\om$. Then there is an $\om$-word $x \in \Sio$ such that $x \notin L(\mathcal{M}_z)$. But 
$L(\mathcal{M}_{\theta(z)}) = (L  \otimes   \Sio)  \cup     (\Sio    \otimes   L(\mathcal{M}_{z}))$ thus 
$\{ \sigma \in \Sio \mid \sigma \otimes x \in L(\mathcal{M}_{\theta(z)}) \} = L$ is a ${\bf \Si}^1_1$-complete set. 
The function $\psi_x : \sigma \ra \sigma \otimes x$ is continuous. 
This implies that $L(\mathcal{M}_{\theta(z)}) $ 
is not a Borel set because otherwise $L= \{ \sigma \in \Sio \mid \sigma \otimes  x \in L(\mathcal{M}_{\theta(z)}) \}= \psi_x^{-1}(L(\mathcal{M}_{\theta(z)}))$
 would be also Borel as the inverse image of a Borel set by a continuous function \cite{Kechris94}. 
\nl
Then it is easy to see that 
 $L^B(\mathcal{T}_{H \circ \theta (z)}) = L(\mathcal{M}_{ \theta (z)})^a \cup  [ \Si^{\om, \om}- (\Sio)^a ] $ 
is not a Borel set. 
But it was proved in \cite[Lemma 5.2]{Fin04} that every E-recognized language of infinite pictures is a 
${\bf \Si}^0_2$-set and in \cite[Lemma 5.3]{Fin04} that every A-recognized language of infinite pictures is a 
closed set. 
\nl Thus in that case the $\om$-picture language $L^B(\mathcal{T}_{H \circ \theta (z)}) $ is  neither E-recognizable nor A-recognizable.

\hs Finally, using the reduction $H \circ \theta $, we have proved that : 
 $$\{ z \in \mathbb{N} \mid  L(\mathcal{M}_z) = \Si^\om \} \leq_1 \{ z \in \mathbb{N} \mid  L^B(\mathcal{T}_z) \mbox{ is  E-recognizable  }  \}$$ 
$$\{ z \in \mathbb{N} \mid  L(\mathcal{M}_z) = \Si^\om \} \leq_1 \{ z \in \mathbb{N} \mid  L^B(\mathcal{T}_z) \mbox{ is  A-recognizable  }  \}$$ 

\noi and this ends the proof. 
\ep 

\hs  As in \cite{Fin04} we are going to infer from the proof of (high)  undecidability of E-recognizability (respectively, A-recognizability)  some 
other  (high) undecidability results. 

\hs It was proved in \cite{Fin04} that  for any Borel class ${\bf \Si_{\alpha}^0}$ or   ${\bf \Pi_{\alpha}^0}$,   it 
is undecidable whether a given B\"uchi-recognizable language of $\om$-pictures is in 
${\bf \Si_{\alpha}^0}$ (respectively, is in   ${\bf \Pi_{\alpha}^0}$, is a Borel set).
We can deduce from the above proof that the topological complexity of recognizable languages of infinite pictures  is  in fact 
highly undecidable.

\begin{The}
\noi Let $\alpha$ be a non-null countable ordinal. Then  
\begin{enumerate}
\ite $ \{  z \in \mathbb{N}  \mid   L^B(\mathcal{T}_z) \mbox{ is in the Borel class } {\bf \Si}^0_\alpha \}$ is  $\Pi_2^1$-hard. 
\ite  $ \{  z \in \mathbb{N}  \mid   L^B(\mathcal{T}_z)  \mbox{ is in the Borel class } {\bf \Pi}^0_\alpha \}$ is  $\Pi_2^1$-hard. 
\ite  $ \{  z \in \mathbb{N}  \mid  L^B(\mathcal{T}_z)  \mbox{ is a  Borel set } \}$ is  $\Pi_2^1$-hard. 
\end{enumerate}
\end{The}

\proo  We can use the same reduction $H \circ \theta $ as in the proof of Theorem \ref{EA}. We have seen that there are two cases. 
\nl {\bf First case.}  $L(\mathcal{M}_z)= \Si^\om$. Then $L(\mathcal{M}_{\theta(z)}) = \Si^\om$ and 
$L^B(\mathcal{T}_{H \circ \theta (z)}) =  \Si^{\om, \om}$. In particular $L^B(\mathcal{T}_{H \circ \theta (z)}) =  \Si^{\om, \om}$
 is an open and closed subset of $\Si^{\om, \om}$
and it belongs to all Borel classes $ {\bf \Si}^0_\alpha$ and  ${\bf \Pi}^0_\alpha$. 
\nl {\bf Second case.} $L(\mathcal{M}_z) \neq  \Si^\om$. Then we have seen that $L^B(\mathcal{T}_{H \circ \theta (z)}) $
 is not a Borel set. 
\nl  Finally, using the reduction $H \circ \theta $, the result follows from the $\Pi_2^1$-completeness of the universality problem for 
$\om$-languages of Turing machines. 
\ep 

\hs We now come to the complementability  problem. The class of      B\"uchi-recognizable  languages of   infinite pictures 
is not closed under complement \cite{ATW02}. 
Thus the question naturally arises: ``can we decide whether the complement of a  B\"uchi-recognizable   language of  infinite pictures is B\"uchi-recognizable?". 
It has been proved in \cite{Fin04} that this problem is undecidable. We are going to prove that it is in fact $\Pi_2^1$-complete. 

\hs Another classical problem is the determinizability problem: ``can we decide whether a given  recognizable   language of  infinite pictures is recognized by a 
deterministic tiling system?". Again this problem has been proved to be undecidable in  \cite{Fin04} and we shall prove it is in fact $\Pi_2^1$-complete. 
\nl  Recall  that
 a tiling system is called deterministic if on any picture 
it allows at most one tile covering the origin, the state assigned to position 
$(i+1, j+1)$ is uniquely determined by the states at positions $(i, j), (i+1, j), (i, j+1)$ 
and the states at the border positions $(0, j+1)$ and $(i+1, 0)$ are determined by the state 
$(0, j)$, respectively $(i, 0)$, \cite{ATW02}. 
\nl As remarked in \cite{ATW02}, the hierarchy proofs of the classical 
Landweber hierarchy defined using deterministic $\om$-automata ``carry over without essential 
changes to pictures". In particular it is easy to see that a language of $\om$-pictures which 
is B\"uchi-recognized by a \de  tiling system is a ${\bf \Pi}^0_2$-set and that a language of $\om$-pictures which 
is Muller-recognized by a \de  tiling system is a boolean combination of ${\bf \Pi}^0_2$-sets, hence a ${\bf \Delta}^0_3$-set. 

\hs We can now state the following results.  

\begin{The}
\noi  The determinizability problem and the complementability  problem for B\"uchi-recognizable languages of infinite pictures 
are   $\Pi_2^1$-complete,  i.e. : 
\begin{enumerate}
\ite 
$\{  z \in \mathbb{N}  \mid  
L^B(\mathcal{T}_z) \mbox{ is B\"uchi-recognizable  by a deterministic tiling system}\}$ is $\Pi_2^1$-complete. 
\ite
$\{  z \in \mathbb{N}  \mid 
 L^B(\mathcal{T}_z) \mbox{ is Muller-recognizable  by a deterministic tiling system }\}$ is $\Pi_2^1$-complete. 
\ite
 $\{  z \in \mathbb{N}  \mid  \exists y ~~ \Si^{\om, \om} - L^B(\mathcal{T}_z) =  L^B(\mathcal{T}_y) \}$ is $\Pi_2^1$-complete. 
 \end{enumerate}
\end{The}

\proo  
 It is easy to see that the set  $D$ of indices of  deterministic tiling systems equipped with a B\"uchi acceptance condition  is recursive. The formula 
 $\exists y \in D ~~  L^B(\mathcal{T}_z)=L^B(\mathcal{T}_y)$ can be 
written : ``$\exists y [ y \in D  \mbox{ and  } L^B(\mathcal{T}_z)=L^B(\mathcal{T}_y) ] $" and it can be expressed by a $\Pi_2^1$-formula because the 
quantifier $\exists y $ is of  type $0$ and ``$L^B(\mathcal{T}_z)=L^B(\mathcal{T}_y)$" can be expressed by a $\Pi_2^1$-formula. Thus  
the set  $\{  z \in \mathbb{N}  \mid  \exists y \in D ~~  L^B(\mathcal{T}_z)=L^B(\mathcal{T}_y) \}$ is in the class  $\Pi_2^1$. 
\nl The case of  deterministic tiling systems with {\it Muller} acceptance condition is very similar. Details are here left to the reader. 
\nl On the other hand  ``$\Si^{\om, \om} - L^B(\mathcal{T}_z) =  L^B(\mathcal{T}_y)$"
 can be expressed by a $\Pi_2^1$-formula so  `` $\exists y ~~ \Si^{\om, \om} - L^B(\mathcal{T}_z) =  L^B(\mathcal{T}_y)$" can be 
expressed by a $\Pi_2^1$-formula because     the quantifier $\exists y$ is of type $0$. Thus the set 
 $\{  z \in \mathbb{N}  \mid  \exists y ~~ \Si^{\om, \om} - L^B(\mathcal{T}_z) =  L^B(\mathcal{T}_y)   \}$ 
 is in the class $\Pi_2^1$. 

\hs To prove completeness, we use the same reduction $H \circ \theta $ as in the proof of Theorem \ref{EA}. We have seen that there are two cases. 
\nl {\bf First case.}  $L(\mathcal{M}_z)= \Si^\om$ and then  
$L^B(\mathcal{T}_{H \circ \theta (z)}) =  \Si^{\om, \om}$. In particular $L^B(\mathcal{T}_{H \circ \theta (z)}) =  \Si^{\om, \om}$
is accepted by a B\"uchi deterministic tiling system and also by a Muller deterministic tiling system. Morever its complement is empty so it is 
B\"uchi (or Muller) recognized by a   tiling system. 
\nl {\bf Second case.} $L(\mathcal{M}_z) \neq  \Si^\om$. Then we have seen that $L^B(\mathcal{T}_{H \circ \theta (z)}) $
 is not a Borel set. Thus in that case $L^B(\mathcal{T}_{H \circ \theta (z)}) $ cannot be accepted by any deterministic tiling system with 
B\"uchi or Muller acceptance condition. 
 Moreover  its complement $\Si^{\om, \om} - L^B(\mathcal{T}_{H \circ \theta (z)})$ is  not a 
 ${\bf \Si}^1_1$-subset of  $\Si^{\om, \om}$ because otherwise $L^B(\mathcal{T}_{H \circ \theta (z)})$ would be in 
${\bf \Delta}^1_1={\bf \Si}^1_1\cap {\bf \Pi}^1_1$ which is the class of Borel sets by Suslin's Theorem. Thus 
$\Si^{\om, \om} - L^B(\mathcal{T}_{H \circ \theta (z)})$  cannot be 
B\"uchi-recognizable because it is not a ${\bf \Si}^1_1$-subset of  $\Si^{\om, \om}$ and 
$TS(\Si^{\om, \om}) \subseteq  \Si^1_1 \subseteq {\bf \Si}^1_1$, see \cite{ATW02}. 
\nl  Finally, using the reduction $H \circ \theta $, the result follows from the $\Pi_2^1$-completeness of the universality problem for 
$\om$-languages of Turing machines. 
\ep 

\hs
We gave in \cite{Fin04}  a solution to a question of \cite{ATW02}, showing 
that all languages of infinite pictures which are accepted row by row 
by B\"uchi or Choueka 
automata reading words of 
length $\om^2$ are B\"uchi recognized by a finite tiling system, but 
the converse is not true. Then we showed that one cannot decide whether a given B\"uchi-recognizable language of infinite pictures is 
accepted row by row by a  B\"uchi or Choueka 
automaton reading words of 
length $\om^2$. We are going to show now that this decision problem is actually also  $\Pi_2^1$-complete. 

\hs Recall that an $\om^2$-word $x$
 over the alphabet $\Si$ is a
sequence of length $\om^2$ of letters in $\Si$. It is denoted by 
$(x(i))_{0\leq i<\om^2}=x(0).x(1).x(2)\ldots x(i) \ldots$ ,  
where for all $i$, $0\leq i<\om^2$, $x(i)$ is a letter in $\Si$. 
\nl  The set of  $\om^2$-words over 
 $\Si$ is denoted by $\Si^{\om^2}$. An  $\om^2$-language over $\Si$ is a subset of $\Si^{\om^2}$.
\nl To define a notion of acceptance row by row of an $\om$-picture we first  associate, 
to an infinite picture $p  \in  \Si^{\om, \om}$,   an $\om^2$-word 
$\bar{p} \in \Si^{\om^2}$ which is defined by $\bar{p}(\om.n + m)=p(m+1, n+1)$ 
for all integers $n, m \geq 0$. 
\nl This can be extended to languages of infinite pictures: for $L \subseteq \Si^{\om, \om}$ 
we denote $\bar{L}=\{\bar{p} \mid p \in  L \}$ so $\bar{L}$ 
 is an $\om^2$-language over $\Si$.

\hs We refer the reader to \cite{Fin04} for a precise definition of generalized B\"uchi automaton acceptings words of ordinal length. 
We recall now the following  definition.

\begin{Deff}
A language  of infinite pictures $L \subseteq \Si^{\om, \om}$ 
is accepted row by row by an ordinal B\"uchi 
automaton  if and only if 
the  $\om^2$-language $\bar{L}$ is regular, i.e. is  accepted by an ordinal  B\"uchi automaton. 
We denote $BA(\Si^{\om, \om})$ the set of languages $L \in \Si^{\om, \om}$ such that $\bar{L}$ is regular. 
\end{Deff}

\noi We can now state the following result. 

\begin{The}
\noi  The  problem to determine whether a given 
 B\"uchi-recognizable language of infinite pictures is accepted row by row by an ordinal B\"uchi 
automaton, is $\Pi_2^1$-complete,  i.e. : 
$$ \{  z \in \mathbb{N}  \mid  
\bar{L}^B(\mathcal{T}_z) \mbox{ is regular } \} \mbox{ is } \Pi_2^1\mbox{-complete}.$$
\end{The}

\proo   Recall that, for each language of infinite pictures which is accepted row by row 
by a B\"uchi  automaton reading words of 
length $\om^2$, it was constructed in \cite{Fin04} a Muller tiling system accepting it. Then, using \cite[Theorem 1]{ATW02}, one can effectively construct
a B\"uchi tiling system accepting the same language. 
The set $T_R$  of indices  of B\"uchi tiling systems constructed from the proof of  \cite[Theorem 4.1]{Fin04} and \cite[Theorem 1]{ATW02}
  is easily seen to be recursive. Notice  that $T_R$ does not contain all  
indices of B\"uchi tiling systems accepting languages  in $BA(\Si^{\om, \om})$. But for each language $L$  in $BA(\Si^{\om, \om})$ there 
is an index $z \in T_R$ such that $L=L^B(\mathcal{T}_z)$. 
\nl We can then express ``$\bar{L}^B(\mathcal{T}_z) \mbox{ is regular }$" by the formula 
``$\exists y [ (~y \in T_R ~ )$ and $L^B(\mathcal{T}_z)=L^B(\mathcal{T}_y)$".  This is a $ \Pi_2^1$-formula because
 ``$L^B(\mathcal{T}_z)=L^B(\mathcal{T}_y)$" can be expressed by a $ \Pi_2^1$-formula and the quantifier $\exists y $ is of type $0$. 

\hs To prove completeness we can use the same reduction $H \circ \theta $ as in the proof of Theorem \ref{EA}. We have seen that there are two cases. 
\nl {\bf First case.}  $L(\mathcal{M}_z)= \Si^\om$ and then  
$L^B(\mathcal{T}_{H \circ \theta (z)}) =  \Si^{\om, \om}$. In particular,  $L^B(\mathcal{T}_{H \circ \theta (z)}) =  \Si^{\om, \om}$
is accepted row by row by an ordinal B\"uchi 
automaton. 
\nl {\bf Second case.} $L(\mathcal{M}_z) \neq  \Si^\om$. Then we have seen that $L^B(\mathcal{T}_{H \circ \theta (z)}) $
 is not a Borel set. Thus in that case $\bar{L}^B(\mathcal{T}_{H \circ \theta (z)}) $ is not a regular  $\om^2$-language because 
otherwise $L^B(\mathcal{T}_{H \circ \theta (z)}) $  would  be a Borel set (of rank smaller than or equal to 5), see \cite[Proposition 4.2]{Fin04}. 

\hs  Finally, using the reduction $H \circ \theta $, the completeness result follows from the $\Pi_2^1$-completeness of the universality problem for 
$\om$-languages of Turing machines. 

\ep

\section{Concluding Remarks} 

\noi We have given in this paper the exact degree of numerous natural decision problems for recognizable languages of infinite pictures. 
This way we have given examples of natural highly undecidable problems which are complete at the first or at the second level of the analytical hierarchy. 
Notice that many examples of $\Si_1^1$-complete problems are already known, such as the recurring tiling problem, see for instance 
\cite{Harel83,Harel86,HirstHarel96a}. 
But it seems that very few natural problems, except some problems about $\om$-languages of Turing machines, are known to be $\Pi_2^1$-complete. 
One of the motivation of Castro and Cucker in \cite{cc} was actually to ``give natural complete problems for the 
lowest levels of the analytical hierarchy which constitute an analog of the classical complete problems given in recursion theory for the arithmetical hierarchy".
So we have added in this paper many new examples which complete the work of \cite{cc}. 

\hs
Notice that in another paper we have also given many natural  $\Pi_2^1$-complete problems about the infinite  behaviour of very 
simple finite machines like $1$-counter automata or $2$-tape automata, \cite{Fin-HI}. 

\hs We hope also that our results could  be  useful in other connected areas, for instance in  the study of the infinite behaviour of cellular automata, 
\cite{Wolfram,Delorme-Mazoyer}. 

\hs {\bf Acknowledgements.}
\nl  Thanks to  the anonymous referees 
for useful comments on a preliminary version of this paper.

\end{document}